\documentclass[12pt]{article}
\usepackage{cite} 
\usepackage{amsfonts,latexsym}
\usepackage{epsfig}
\begin{document}
\newcommand{\de}{\delta}\newcommand{\ga}{\gamma}
\newcommand{\e}{\epsilon} \newcommand{\ot}{\otimes}
\newcommand{\be}{\begin{equation}} \newcommand{\ee}{\end{equation}}
\newcommand{\ba}{\begin{array}} \newcommand{\ea}{\end{array}}
\newcommand{\beq}{\begin{equation}}\newcommand{\eeq}{\end{equation}}
\newcommand{\tmod}{{\cal T}}\newcommand{\amod}{{\cal A}}
\newcommand{\bemod}{{\cal B}}\newcommand{\cmod}{{\cal C}}
\newcommand{\dmod}{{\cal D}}\newcommand{\hmod}{{\cal H}}
\newcommand{\s}{\scriptstyle}\newcommand{\tr}{{\rm tr}}
\newcommand{\einsop}{{\bf 1}}
\def\R{\overline{R}} \def\doa{\downarrow}
\def\dag{\dagger}
\def\ve{\epsilon}
\def\si{\sigma}
\def\ga{\gamma}
\def\no{\nonumber}
\def\le{\langle}
\def\re{\rangle}
\def\lt{\left}
\def\rt{\right}
\def\dwn{\downarrow}   
\def\up{\uparrow}
\def\dag{\dagger}
\def\nonum{\nonumber}
\newcommand{\reff}[1]{eq.~(\ref{#1})}

\title{Behaviour of the energy gap in a model of Josephson 
coupled Bose-Einstein condensates }

\author{A. P. Tonel$^{1}$\footnote{sponsored by CNPq-Brazil},
J. Links$^{1}$  
and A. Foerster$^{2}$
\vspace{1.0cm}\\
$^{1}$ Centre for Mathematical Physics, School of Physical Sciences \\ 
The University of Queensland, Queensland, 4072, Australia 
\vspace{0.5cm}\\
$^{2}$ Instituto de F\'{\i}sica da UFRGS \\ 
Av. Bento Gon\c{c}alves 9500, Porto Alegre, RS - Brazil}

\maketitle

\begin{abstract}
In this work we investigate the energy gap between the ground state 
and the first excited state in  a model of two 
single-mode Bose-Einstein condensates 
coupled via Josephson tunneling. The energy gap is never zero 
when the tunneling 
interaction is non-zero. The gap exhibits no local minimum below a 
threshold coupling which separates a delocalised phase 
from a self-trapping phase that occurs 
in the absence of the external potential. Above this
threshold point one minimum occurs close to the Josephson regime, 
and a set of minima and maxima appear
in the Fock regime. Expressions for the position
of these minima and maxima are obtained.  
The connection between these minima and maxima and the dynamics 
for the expectation value of the relative number of particles is 
analysed in detail.
We find that the dynamics of the system changes as the coupling 
crosses these points. 
\end{abstract}


\vfil\eject
%%%%%%%%%%%%%%%%%
\section{Introduction}
Research into Bose-Einstein condensates (BECs) continues to attract 
considerable attention from both theoretical 
and experimental sides due to its contribution to the 
comprehension of physical phenomena 
emerging from mesoscopic systems. Recently, with the  
modernisation of experimental techniques, 
BECs have been produced and investigated by several research groups using 
magnetic traps with different kinds of 
confining geometries \cite{cw,ak} and chip devices \cite{chip1,chip2}.
{}From the theoretical point of view, the model of two single-mode BECs 
coupled via Josephson tunneling (see eq. \ref{ham} in the next section 
and a comprehensive review in \cite{leg}) has been a successful model in 
further understanding the experimental realization of BECs in  dilute 
alkali gases \cite{anderson}. The success of this model is based on two 
main reasons: first is that despite its apparent simplicity, the model
captures the physics behind the macroscopic quantum phenomena that emerge 
from these systems; second is that it is also applicable in other studies in 
mesoscopic solid state Josephson junctions \cite{leg,int8,mss}, non-linear 
optics \cite{saf}, statistical physics of spin systems \cite{bjp,bj} and in 
nuclear physics \cite{lmg,lmg1}. A theoretically important feature of 
this model is it can be formulated through the Quantum Inverse Scattering 
Method establishing solvability by Bethe Ansatz methods \cite{zhou1,zhou2}. 

Many aspects of this model have already been investigated, such as 
the dynamics of the population imbalance \cite{milb,clsx,our}, the 
imbalance fluctuation \cite{sr,our} and entanglement 
\cite{saf,hmm,vpa,our,pd} amongst others \cite{kalo,kalo1,kalo2}.  
However an understanding of the properties of the energy gap as a function of
the external potential, and an investigation of the dynamical
consequences, appears lacking. This study is of interest since it could
feasibly be tested in an experimental setting, particularly in cases where
the external potential is the only tunable interaction, 
as may be expected in solid state systems. 
Here, we will expand on this aspect of the model by undertaking a 
detailed analysis of the 
energy gap behaviour across all coupling regimes.
Specifically we will show that the energy gap function presents no local 
minimum below a threshold coupling, defined below,  
while above this threshold 
point one minimum occurs close to the
Josephson regime and a set of minima and maxima appear in the Fock regime.
The connection between these minima and maxima 
and the dynamics for the relative number of 
particles is investigated. We find that these minima and maxima 
act as a transition point separating entirely different dynamical behaviours. 

The paper is organised in the following way. In section II we present the model and discuss the energy gap 
for different choices of the tunneling coupling and different regimes. 
In section III we discuss the correspondence between the minima and maxima that occur in the energy gap 
function and the temporal evolution for the relative number of particles in different regimes. 
In section IV we compare the quantum results with those obtained from a classical analysis.
Section V  is reserved for our conclusions.

%%%%%%%%%%%%%%%%%%%%%%%%%
\section{Energy gap}
The model we will study, based on the two-mode approximation  
as described in \cite{leg} (see also references therein), is described by the following 
Hamiltonian:
\begin{equation}
H=\frac{k}{8}(N_1-N_2)^2 - \frac{\Delta \mu}{2}(N_1-N_2) - \frac{{\cal E}_J}{2}(a^{\dag}_1 a_2 +a_2^{\dag} a_1)
\label{ham}
\end{equation}
where $a^{\dag}_i, a_i$ ($i=1,2$) denote the single-particle creation and annihilation operators 
associated with two bosonic modes and $N_1=a^{\dag}_1 a_1$ and $N_2=a^{\dag}_2 a_2$ are the corresponding number operators. 
The quantity $N=N_1+N_2$ is the total boson number operator and is conserved.
The coupling $k$ provides the strength of the interaction between the bosons, 
$\Delta \mu$ is the external potential and ${\cal E}_J$ is the coupling for the tunneling. 
We mention that throughout we will take $k>0$. 
The change ${\cal E}_J\rightarrow -{\cal E}_{J}$ corresponds to the unitary 
transformation $a_1 \rightarrow a_1,\,a_2\rightarrow -a_2$, while 
$\Delta \mu \rightarrow -\Delta \mu$ corresponds to $a_1 \leftrightarrow a_2$. 
Therefore we will restrict our analysis to the case of 
${\cal E}_J,\,\Delta\mu\geq 0$. It is useful to divide the parameter 
space into a number of regimes; viz. Rabi ($k/{\cal E}_J < 4/N$),
Josephson ($4/N < k/{\cal E}_J << N$) 
and Fock ($k/{\cal E}_J >> N$). In the Rabi and Josephson regimes, coherent 
superposition of the two condensates is expected 
whereas in the Fock regime the two condensates may be essentially localised. 
Note that we adopt the convention ( cf. \cite{ks}) to specifically define the
boundary between the Rabi and Josephson regimes by the threshold
coupling $k/{\cal E}_J = 4/N$. This threshold was found in \cite{milb}
as the transition coupling between macroscopic self-trapping ($k/{\cal
E}_J > 4/N$) and delocalisation $k/{\cal E}_J < 4/N$) for the case
$\Delta\mu =0$. (see also \cite{our,miche,bes}).
The presence of a non-zero external potential changes the coupling ratio
at which this transition occurs, as we will show. However, it is useful
to still refer to $k/{\cal E}_J = 4/N$ as the threshold coupling,  
irrespective of the 
value of the external potential, for other reasons which we will
discuss.

Our main aim here is to investigate the behaviour of the energy 
gap for the Hamiltonian (\ref{ham}).
In the Fock regime, we can preliminarily study this problem by 
putting the tunneling interaction term of the 
Hamiltonian (\ref{ham}) to zero, i.e. ${\cal E}_J =0$. 
In this case all energy levels can be computed exactly. For the case 
of an even number of particles this yields 
\begin{equation}
E_p = \frac{k}{2} p^2+p\Delta \mu ,\;\;\;\;\;  
p= -\frac{N}{2},-\frac{N}{2}+1,...,-1,0,1,...,\frac{N}{2}-1,\frac{N}{2}.
\label{e2}
\end{equation}
In the absence of the external potential ($\Delta \mu =0$) the 
ground state is unique ($E_0 =0$) and all other energies are double-degenerate, since $E_{-p}=E_p$.
By switching on the external potential $\Delta \mu $, 
these degeneracies are broken. 

Fixing the total boson number $N$ and the interaction strength $k$ and allowing the external 
potential $\Delta \mu$ to vary, expression (\ref{e2}) indicates 
two important facts: i) there are no degenerate energy levels,  
only level crossings; ii) all energies are functions of the value of the
external potential value, 
except for $E_0=0$, and all energies of the type $E_{-p}$ and $E_0$ correspond 
to the ground state 
in some region, defined by the strength of the external potential $\Delta \mu$. 
These regions are determined by the level crossings. Results are depicted in 
Fig. 1 for the 
case $N=6, k=6$.

%\vspace{1.2cm}
\begin{figure}[ht]
\begin{center}
\epsfig{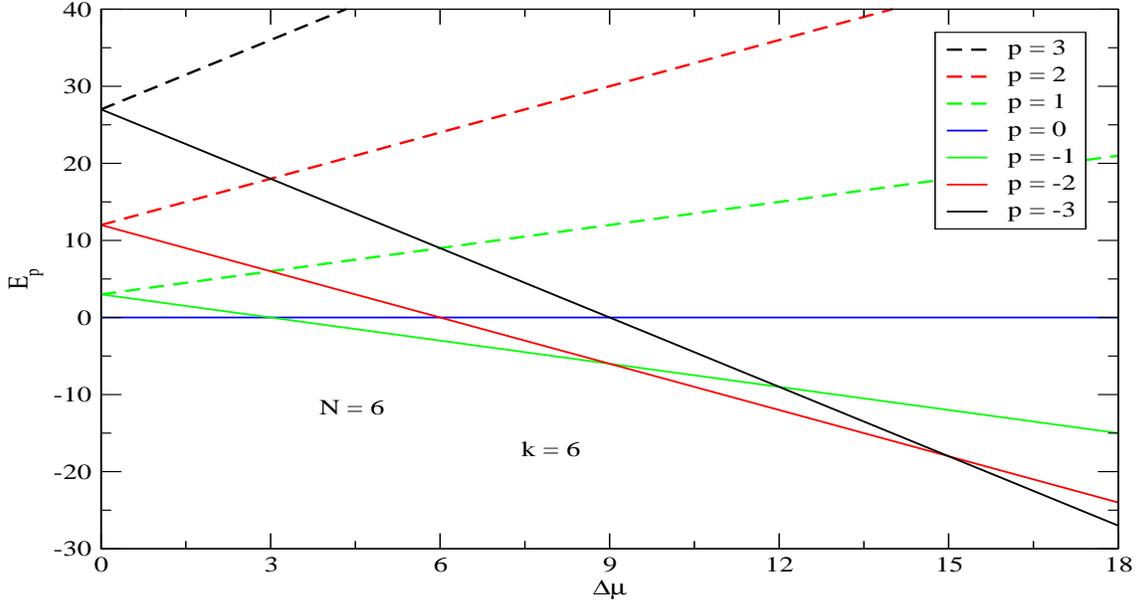}
\caption{Energy levels versus the external potential for $N=6$ and $k=6$. 
The level crossings 
involving the lowest energy level represent changes in the ground state 
structure. The energy gap is also determined by level crossings involving 
the first excited state.}
\label{fig50}
\end{center}
\end{figure}

The energy gap, in each region, is given by
\begin{eqnarray}
\label{neven5}
\Delta_{(j+1,j)}&=&E_{-j-1}-E_{-j}\;\;,\;\;\;\;\;\;\;\;\;\;\;\Delta\mu\in [kj,k(j+1/2)] \nonumber\\
\Delta_{(j-1,j)}&=&E_{-j+1}-E_{-j}\;\;,\;\;\;\;\;\;\;\;\;\;\;\Delta\mu\in [k(j-1/2),kj] \\
\Delta_{(N/2-1,N/2)}&=&E_{-(N/2-1)}-E_{-N/2}\;\;,\;\;\Delta\mu > k(N-1)/2. \nonumber
\end{eqnarray}
Here $\Delta _{(i,j)}$  ($\Delta _{(j,i)}$) denotes the energy gap before (after) the level crossing between the 
energy levels $E_{-i}$ and $E_{-j}$. Notice that at each different region there is either a different 
ground state or a different first excitation. A similar structure has also been found in 
\cite{hosh} by fixing all coupling parameters and varying 
the total number of particles.

Considering the case of an odd total number of particles $N$, eq. (\ref{e2}) is still correct, 
however we have to eliminate the energy level $E_0$ and change the allowed values of the index $p$ to
$p= -N/2,-N/2-1,...,-1/2,1/2,...,N/2-1,N/2$.
Here, at $\Delta \mu =0$, all energies are two-fold degenerate. 
By switching on the external potential $\Delta \mu$, all degeneracies are again broken and   
level crossings appear. 
%The energy gap, in each region, is now given by
%\begin{eqnarray}
%\label{nodd}
%\Delta_{(+1/2,1/2)}&=&E_{1/2}-E_{-1/2}\;\;,\;\;\Delta\mu\in [0,k/2]\nonumber\\
%\Delta_{(3/2,1/2)}&=&E_{-3/2}-E_{-1/2}\;\;,\;\;\Delta\mu\in [k/2,k]\nonumber\\
%\Delta_{(1/2,3/2)}&=&E_{-1/2}-E_{-3/2}\;\;,\;\;\Delta\mu\in [k,3k/2]\nonumber\\
%\Delta_{(5/2,3/2)}&=&E_{-5/2}-E_{-3/2}\;\;,\;\;\Delta\mu\in [3k/2,2k]\nonumber\\\Delta_{(3/2,5/2)}&=&E_{-3/2}-E_{-5/2}\;\;,%\;\;\Delta\mu\in [2k,5k/2]\nonumber\\&.& \\  
%&.&   \nonumber\\
%&.&   \nonumber\\
%\Delta_{(N/2,N/2-1)}&=&E_{-N/2}-E_{-(N/2-1)}\;\;,\;\;\Delta\mu\in [(N-2)k/2,(N-1)k/2] \nonumber\\
%\Delta_{(N/2-1,N/2)}&=&E_{-(N/2-1)}-E_{-N/2}\;\;,\;\;\Delta\mu > (N-1)k/2.\nonumber
%\end{eqnarray}
Comparing the two cases, one can see that  
the gap behaviour of the model with external
potential $\Delta\mu$ and an odd number of particles $N$  
is the same as the profile for a system with $N+1$ particles and
external potential $\Delta\mu+k/2$. 

\begin{figure}[ht]
\begin{center}
\epsfig{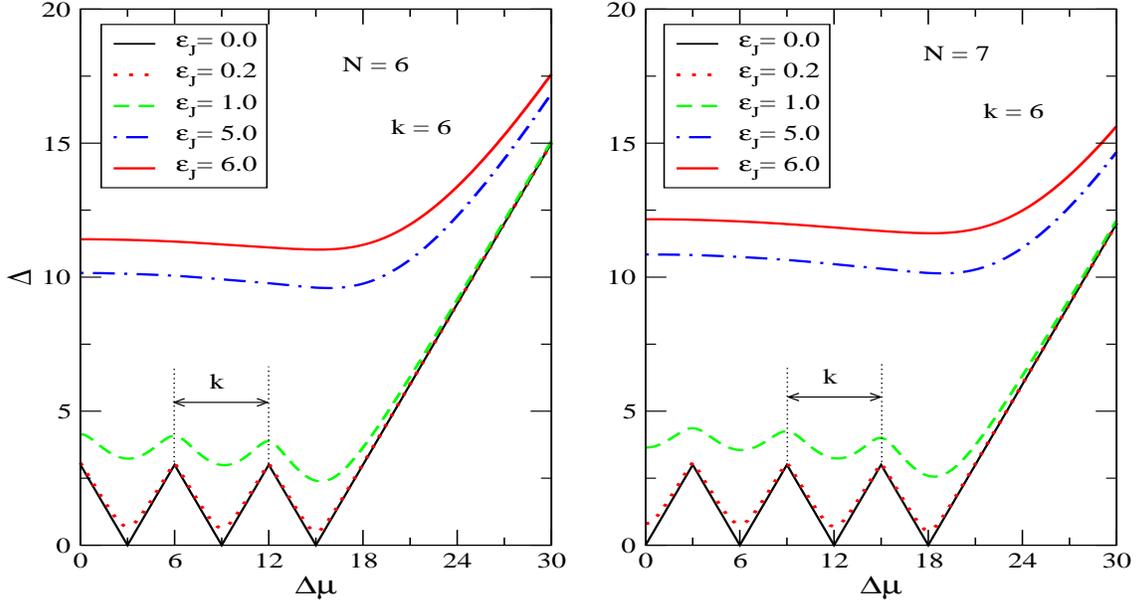}
%\vspace{0.5cm}
\caption{Energy gap $\Delta$ versus the external potential $\Delta \mu$ for 
different choices of the coupling parameter 
${\cal E}_J =0,0.2,1,5,6$. On the left, $N=6$, and on the right, $N=7$. In the extreme Fock regime (${\cal E}_J =0$ ) 
the minima and maxima represent the level crossings as depicted in Fig. 1. 
For small values of ${\cal E}_J $ the difference between two consecutive minima or maxima is constant and equal to $k$. 
As ${\cal E}_J $ increases just one minimum survives.}
\label{fig5}
\end{center}
\end{figure}

Now we consider the situation where the coupling ${\cal E}_J$ between the 
two condensates is switched on
and we study its effect using the numerical diagonalisation of the 
Hamiltonian (\ref{ham}).
In terms of the energy gap, the existence of the threshold coupling
is suggested by the 
curve of the energy gap as 
a function of the external potential.  
Above the threshold ($k/{\cal E}_J>4/N$), one or more local minima
appear 
while no local minimum occurs when $k/{\cal E}_J<4/N$.
Let us first focus our attention on the analysis of the energy gap in the 
region above the 
threshold point, in particular, in the region where $k/{\cal E}_J \geq 1$.
In Fig. \ref{fig5} we plot the energy gap $\Delta$ versus the external 
potential $\Delta \mu $ 
with different values of the coupling parameter ${\cal E}_J$ for the cases $N=6,7$.

\vspace{1.2cm}
We can see clearly the effect produced by the tunneling term ${\cal E}_J$. 
Before switching on this coupling, in the extreme Fock regime $({\cal E}_J =0)$,
the energy gap curve is not differentiable at each minima and maxima.
Here the maxima and minima are determined by level crossing and, consequently, 
either a change in the ground state or the first excited state.  
In this particular situation we can exactly calculate 
the energy gap value in all regions
using equations (\ref{e2}, \ref{neven5}).
It is interesting to observe that at each minimum 
(respectively  maximum) the same value for the energy gap is found.

Turning on the coupling term ${\cal E}_J$ the energy gap function 
becomes smooth. 
An important fact here is that weak tunneling does not 
considerably change the position of the minima and maxima in relation to the ${\cal E}_J=0$ case.
It does change, however, the value of the energy gap at each minimum and maximum.
This effect is strongest in the first region ($ \Delta \mu \in [0,k/2]$), 
reducing gradually at each subsequent region, with the last minimum corresponding to the lowest value of the gap.
Increasing the value of the coupling term ${\cal E}_J$ to take the model out of the Fock regime leads to a suppression in the number of minima and maxima until just one minimum survives.

We now look closer at the set of minima and maxima that appear
in the Fock regime.
The number of these minima and maxima depends on the total number of particles $N$.
If $N$ is even, the number of maxima and minima is the same, $N/2$, while if $N$ 
is odd the number of minima is $(N+1)/2$ and the number of maxima is $(N-1)/2$. 
However, we find that the difference between two consecutive minima (or maxima) is always constant and equal to $k$ 
independent of the total number of particles. 
This allows us to examine a general expression for the position of each minimum 
and maximum of the energy gap that occurs with respect to
the external potential, which we find fits very well with 

\begin{equation}
\Delta \mu_c = \frac{N-l}{2} k+\frac{{\cal E}_J}{N^2} ,\;\;\;\;\;  l= 1,2,3,\dots,N.
\label{rate2}
\end{equation}

Above, odd choices for the index  $l$ ($l=1,3,5,7,...$) generate the minima, while even 
choices ($l=2,4,6,8,...$) the maxima. The particular choice $l=1$ corresponds to 
the lowest minimum, while $l=3$ the second lowest minimum, and so on. 
With the exception of the lowest minimum, all other minima and maxima disappear as the ratio 
${\cal E}_J$ increases. 
This is illustrated in Fig. \ref{fig5}, for ${\cal E}_J=5,6$ 
corresponding to the situation where we approach the Josephson regime after leaving the Fock regime.
For  $k/{\cal E}_J \geq 1$ (Josephson and Fock regimes), 
we find that the position of the minimum can still be determined by the previous expression (\ref{rate2}) using $l=1$.

\begin{figure}[ht]
\begin{center}
\epsfig{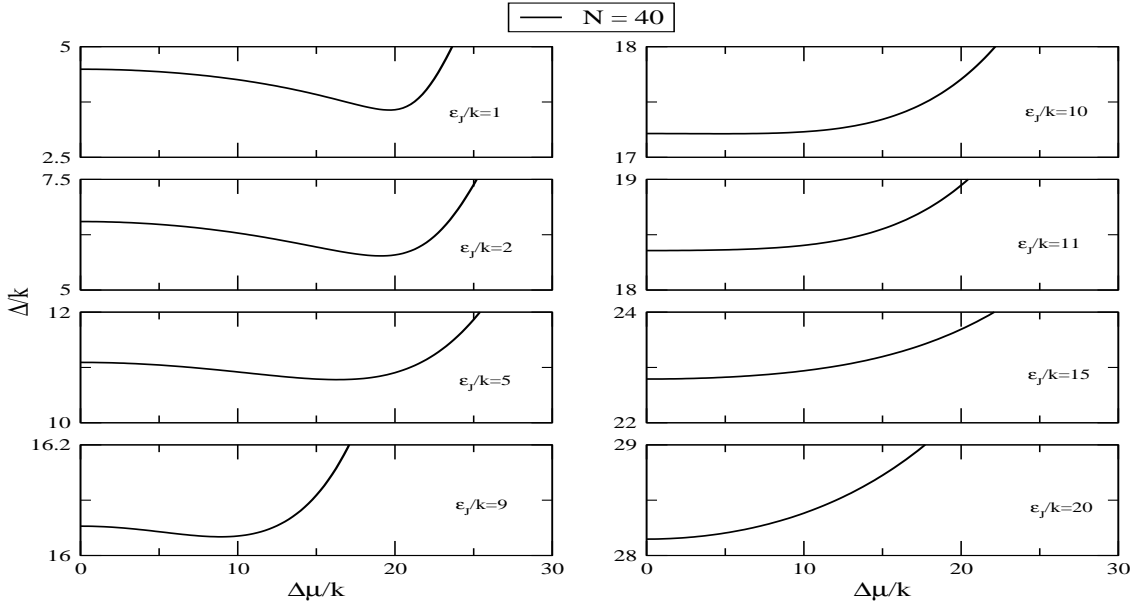}
\caption{Energy gap versus the external potential for $N=40$ and 
different choices of the  
parameter ratio ${\cal E}_J/k =1,2,5,9,10,11,15,20$. On the left (right) 
side we display the points 
below (above) the threshold point ${\cal E}_J/k = 10$. 
On the left hand side we can clearly see 
the existence of a minimal energy gap, while on the right hand side the 
energy gap is 
a monotonic function of the external potential. }
\label{threshold}
\end{center}
\end{figure}

In the interval $k/{\cal E}_J \in (4/N,1]$ 
the energy gap function still has a local minimum, however the position 
at which this occurs is not longer given by (\ref{rate2}) and we have 
not been able to find a simple expression for it. 
The value of $\Delta\mu$ at which the minimum occurs approaches 
zero as the ratio $k/{\cal E}_J$ approaches the threshold coupling. 
Further decreasing the ratio $k/{\cal E}_J$ past the threshold point toward
the Rabi regime, we find a new scenario: 
the energy gap is now a monotonic function for all values of $\Delta \mu$.
This behaviour is illustrated in
Fig. \ref{threshold}, where we plot the energy gap function 
versus the external potential $\Delta \mu$ 
in the region
$k/{\cal E}_J \leq 1$ (Josephson and Rabi regimes, including the 
threshold point) for $N=40$ particles.
In these cases differences between even and odd particle numbers are not
significant as the strength of the tunneling term dominates any 
parity effects due the external potential. 

\vspace{1.0cm}

We emphasise that (\ref{rate2}) was obtained by numerical fitting. It is 
useful to see how this result compares with the predictions of 
perturbation theory. 
It was argued in \cite{zhou1} that for the  
weak tunneling regime perturbation results are only valid for the Fock regime ${\cal E}_J<<k/N$.  
A standard calculation using degenerate perturbation theory shows 
that to first order there is no correction to the values of the applied potential 
which give the minimal energy gaps; i.e. 
$$ 
\Delta \mu_{\rm{pert}}= \frac{N-l}{2}k,~~~l=1,3,5,7,\dots  
$$  
To make comparison between (\ref{rate2}) and
this perturbative result we note 
that for the Fock regime
\begin{eqnarray*} 
\frac{N-l}{2}k &<& \Delta \mu_c  \\
&=& \frac{N-l}{2}k +\frac{{\cal E}_J}{N^2}  \\
&<& \frac{N-l}{2}k +\frac{k}{N^3}  \\
&=&  \left(\frac{N-l}{2} +\frac{1}{N^3}\right)k,   
\end{eqnarray*}
i.e. 
$$
\frac{N-l}{2}k < \Delta \mu_c < \left(\frac{N-l}{2} +\frac{1}{N^3}\right)k . 
$$ 
Thus the formula (\ref{rate2}) agrees with the perturbative result up to a small correction 
of order $N^{-3}$. It is important to stress however our numerical result 
(\ref{rate2}) applies also in the Josephson regime where the perturbative result is not valid.

It was also argued in \cite{zhou1} that perturbation results 
for strong tunneling are only valid for ${\cal E}_J>>kN$. Our numerical results 
show that the existence of minima for the gap 
only occurs for couplings ${\cal E}_J < kN/4$, which does not intersect
the perturbative strong tunneling regime.

%%%%%%%%%%%%%%%%%%%%%%%%%%%%%%%%%%%%%%%%%%%%

In the next section, we will discuss the influence of the gap 
on the quantum dynamics for the relative number of particles.
In particular, we will show that the choice of couplings corresponding to 
the minimal and maximal points of the energy gap 
represent transition points separating  different types of dynamical behaviour.
Fortunately, we can identify all these points using eq. (\ref{rate2}) 
above.

\section{Quantum dynamics}
\begin{figure}[ht]
\begin{center}
\epsfig{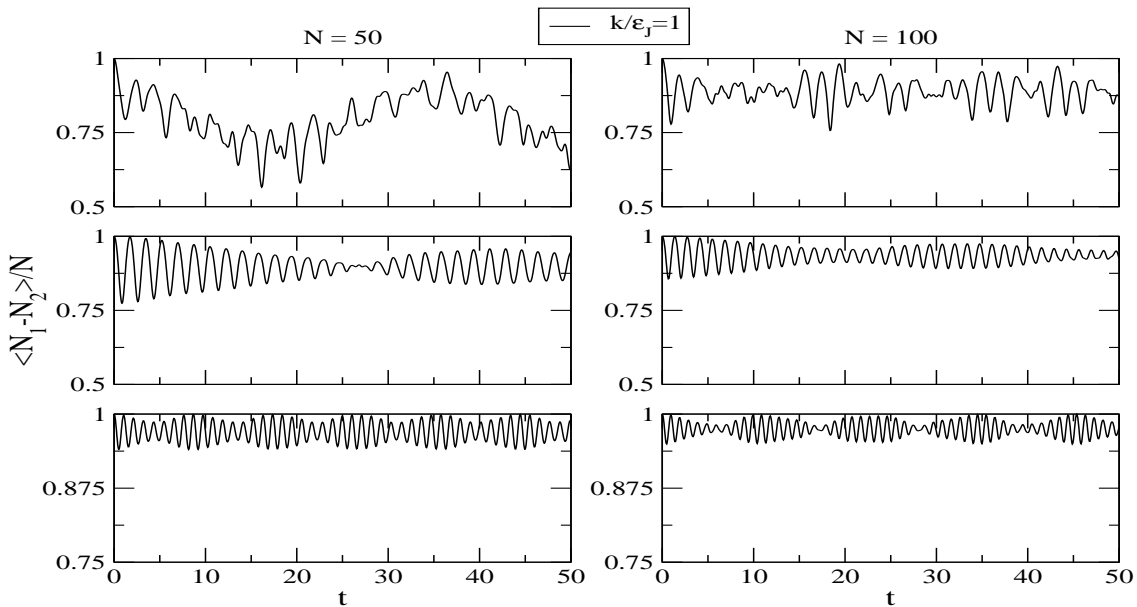}
%\vspace{0.5cm}
\caption{Time evolution of the expectation value of the relative number of particles in the Josephson 
regime $k/{\cal E}_J=1$ (${\cal E}_J=1$), for $N=50,100$. The initial state 
is $|N,0\rangle $. The central graph corresponds to the configuration where the energy gap is a minimum, 
at $\Delta\mu_c/{\cal E}_J=24.5,\,49.5$. 
The other graphs represent points before and after this minimum point  
given by $\Delta\mu/{\cal E}_J= \Delta\mu_c/{\cal E}_J\mp5$.}
\label{fig6}
\end{center}
\end{figure}

We will investigate the quantum dynamics using the same methods we 
employed in \cite{our}. The time evolution of any state is determined 
by $|\Psi(t) \rangle = U(t)|\phi_0 \rangle$, 
where $U$ is the temporal evolution operator given by 
$U(t)=\sum_{m=0}^{M}|m \rangle \langle m|\exp(-i E_m t)$,  
$|m\rangle$ is an eigenstate with energy $E_m$ and $|\phi_0 \rangle$ 
represents the initial state.
Using these expressions we can compute the expectation value of the 
relative number of particles
\begin{equation}
\langle(N_1-N_2)(t)\rangle=\langle \Psi (t)|N_1-N_2|\Psi (t)\rangle. 
\end{equation} 
In Fig. \ref{fig6} we plot the expectation value for the 
relative number of particles in the Josephson regime 
($k/{\cal E}_J=1$) for different ratios of the coupling $\Delta\mu/{\cal E}_J$
for $N=50,100$ and with the initial state $|N,0\rangle$. 

\vspace{1.2cm}

The central graph in each column represents the dynamics when the 
system is in the 
configuration where the energy gap is minimal, 
which can be calculated through eq. (\ref{rate2}). 
This corresponds to $\Delta\mu_c/{\cal E}_J=24.5,\,49.5$ 
for the cases $N=50,\,100$, respectively.
The other graphs in each column, above and below, depict the dynamics for  
cases where $\Delta\mu/{\cal E}_J= \Delta\mu_c/{\cal E}_J\mp5$ in 
Fig. \ref{fig6}.
The coupling corresponding to the minimal energy gap  
signifies a point where 
the dynamics changes drastically: 
before it the behaviour is clearly non-periodic
(the first box in each column), while after it the oscillations  
follow a collapse and revival sequence (the last box in each column). 
A noticible feature here is that in the same region, 
both graphs $N=50,\,100$ show basically the 
same qualitative behaviour. The collapse and revival pattern, 
however, is more pronounced when the 
number of particles is larger. 
Further increasing the ratio $\Delta\mu/{\cal E}_J$ 
beyond the value corresponding to the minimal energy gap, 
the behaviour is still of a collapse and revival type, however 
the amplitude of oscillation and the collapse and revival time both decrease.
   
\begin{figure}[ht]
\begin{center}
\epsfig{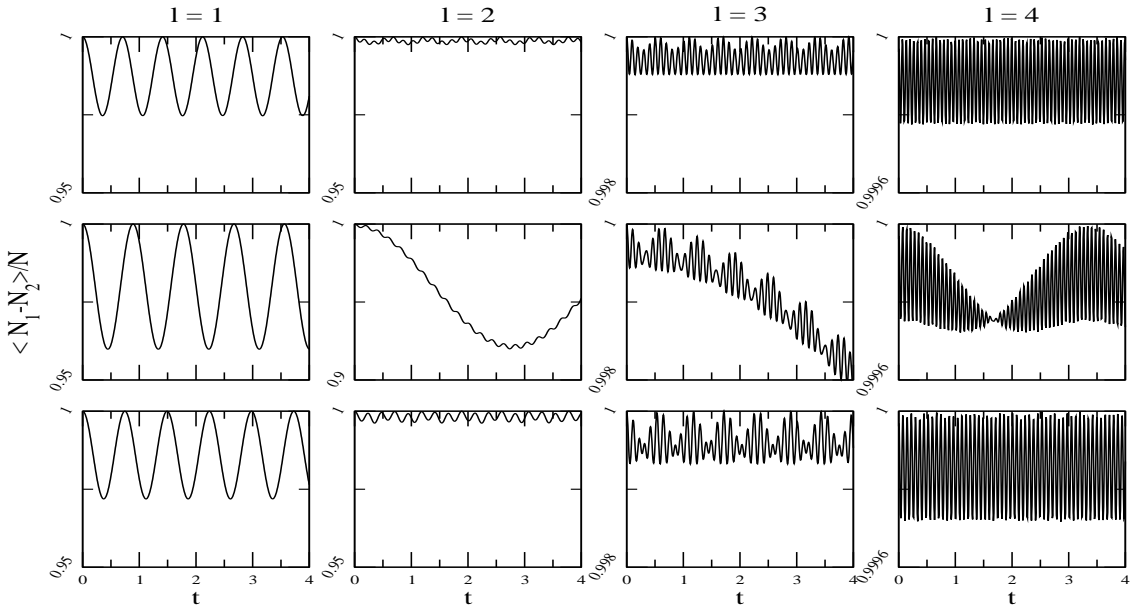}
\caption{Dynamics of the expectation value of the relative number of particles for $N=50$, $k/{\cal E}_J=60$ (${\cal E}_J=1$) and 
with the initial 
state $|N,0\rangle $. The central graph, at each column, represents the point where the energy gap is 
minimal ($l=1,3$) or maximal ($l=2,4$), occurring at 
$ \Delta\mu_c/{\cal E}_J =1470,1440,1410,1380$ (from left to right).
In each case the graphs above and below  show dynamics of points 
around the minimum or maximum, 
chosen as  $\Delta\mu/{\cal E}_J =\Delta\mu_c/{\cal E}_J \mp 5$. 
It is clear that each minimum  and maximum signifies a change in 
the dynamical behaviour.}
\label{fig7}
\end{center}
\end{figure}

Next we turn our attention to study the connection between 
the dynamics and the energy gap for
the crossover region between the Josephson and Fock regimes, where a
set of minima and maxima in the energy gap appear as a function 
of the external potential. 
Again, the position of these minima and maxima can be 
obtained from eq. (\ref{rate2}). 
In Fig. \ref{fig7} we plot the temporal evolution of the 
expectation value for the relative 
number of particles for the case $N=50$
for the coupling ratio $k/{\cal E}_J=60$, 
using the initial state $|N,0\rangle$.
The central graph in each column represents the point where the energy gap is 
minimal ($l=1,3$) or maximal ($l=2,4$), occurring at 
$\Delta\mu_c/{\cal E}_J=1470,\,1440,\,1410,\,1380$ (from left to right).
The other graphs above (respectively below) in each column represent 
the dynamics for the 
situation where $\Delta\mu$ is smaller (respectively larger) than   
$\Delta\mu_c$.
The first and third columns represent the 
dynamics for cases where the energy gap is near the two lowest minima, while
the second and fourth represent the behavior near the two lowest maxima.

\vspace{1.2cm}
The minima and maxima for the energy gaps represent 
frontiers between different types 
of oscillations in the neighbourhood of these points. 
After the last minimum at $l=1$ the expectation value evolves periodically, 
as in the last graph of the first column. 
However, the amplitude of the oscillations becomes smaller as the 
ratio $\Delta\mu/{\cal E}_J$ becomes larger.

\begin{figure}[ht]
\begin{center}
\epsfig{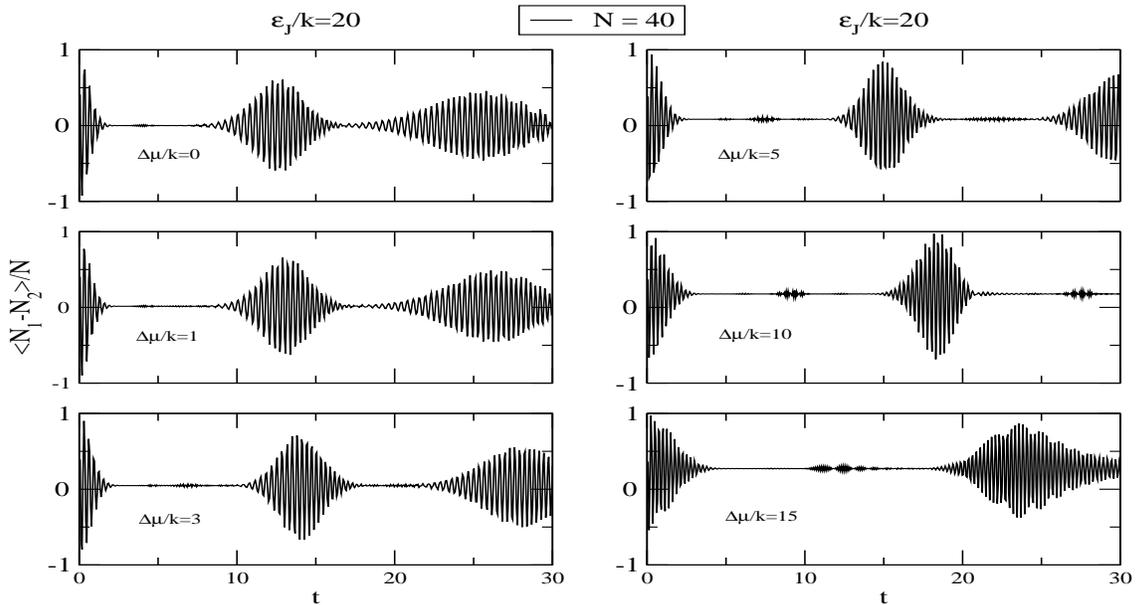}
\caption{Expectation value for the relative number of particles versus 
time for different values of the 
external potential in the regime between Josephson and Rabi 
(below the threshold coupling). Here,, we are fixing ${\cal E}_J=20$ and $k=1$
The initial state is $|N,0\rangle $ with $N=40$. Increasing the external 
potential produces increasing 
collapse and revival time and also leads to a self-trapping phase.}
\label{n40pq}
\end{center}
\end{figure}

Finally we investigate the dynamics for the situation where the system 
is below the threshold coupling ($k/{\cal E}_J < 4/N$) 
where the energy gap does not exhibit a local minimum, as already mentioned. 
In Fig. \ref{n40pq} we plot the expectation value for the 
relative number of particles for 
different choices of the ratio ${\cal E}_J/k $ in the regime 
between Josephson and Rabi 
for $N=40$ and using the initial state $|N,0\rangle$. 
Here the dynamics displays  collapse and revival of oscillations 
for any value of the external potential.  
The presence of an external potential increases the 
collapse and revival time, which tends to 
infinity as the external potential approaches infinity.  
Moreover, increasing the external potential also  
leads the system out of a delocalised phase into one of self-trapping. 
In fact, it is apparent that for any choice of ${\cal E}_J/k $ one can
choose a suitably large $\Delta\mu$ which will dominate the dynamics and
the system will necessarily display self-trapping. 

\vspace{1.0cm}

%%%%%%%%%%%
\section{Classical dynamics} 

Here we compare the above results with those that are obtained by an analogous classical 
analysis of the model. Defining $z=(N_1-N_2)/N$  we may equivalently consider the Hamiltonian 
\cite{ks,rsfs}
\begin{equation}
H(z,\phi)= \frac{{\cal E}_JN}{2} \left(\frac{\lambda}{2}z^2-\beta z - \sqrt{1-z^2}\cos({2\phi}/{N})\right)
\label{hsc}
\end{equation}
where $$\lambda=\frac{kN}{2{\cal E}_J},~~~~~\beta=\frac{\Delta\mu}{{\cal E}_J} $$
and $z,\,\phi$ are canonically conjugate variables.
The Hamiltonian (\ref{hsc}) obeys the symmetries 
 \begin{eqnarray*}\left.H\left(z,\phi\right)\right|_{\lambda,\beta}
&=&-\left.H\left(z,\phi+N\pi/2\right)\right|_{-\lambda,-\beta}\\
 \left.H\left(z,\phi\right)\right|_{\lambda,\beta}
&=&\left.H\left(-z,\phi\right)\right|_{\lambda,-\beta}. 
\end{eqnarray*}
In the usual way the dynamics is given by Hamilton's equations
\begin{eqnarray}
\dot{\phi}&=& \frac{\partial H}{\partial z}=\frac{{\cal E}_JN}{2} \left(
\lambda z - \beta +\frac{z}{\sqrt{1-z^2}}\cos({2\phi}/{N})\right) \\
\dot{z}&=&-\frac{\partial H}{\partial \phi}=-{\cal E}_J 
\left(\sqrt{1-z^2}\sin({2\phi}/{N})\right).
\end{eqnarray}

The classical dynamical evolution of the system is constrained to the level curves of $H$ in phase space. 
The nature of the level curves is largely determined by the fixed points of $H$ determined by 
$\dot{z}=\dot{\phi}=0$, and transitions in 
the dynamical behaviour can be identified with fixed point bifurcations. We find the following classification
of the fixed points: 

\begin{itemize}
\item $\phi=0$ and $z$ is a solution of 

\begin{equation}
\lambda z = - \frac{z}{\sqrt{1-z^2}}+\beta
\label{teq}
\end{equation}
which has a unique real solution for $\lambda\geq 0$. 
For this solution the Hamiltonian attains a local minimum.

\item $\phi={N\pi}/{2}$  and $z$ is a solution of 
\begin{equation}
\lambda z = \frac{z}{\sqrt{1-z^2}}+\beta.
\label{teq1}
\end{equation}
This equation has either one or three real solutions. \\

a) $ 0 \leq \lambda<1$ \\
\noindent For any value of $\beta$ there is just one real solution, for which the Hamiltonian 
attains a local maximum. \\

b) $\lambda >1$ \\
\noindent Here a transition value for $\beta_{c}$ appears, which is dependent on $\lambda$. 
For $\beta<\beta_{c}$, the equation has two locally maximal solutions and one saddle point, 
while for $\beta>\beta_{c}$ the equation has just one real solution, a locally 
maximal fixed point. Thus a fixed point bifurcation occurs at $\beta_c$.

\end{itemize}

We remark that in the case $\lambda=1$ the transition value is given by 
$\beta_c=0$. Thus in the absence of the external potential the classical dynamics predicts 
a transition coupling of $k/{\cal E}_J=2/N$ \cite{ks}, whereas for the quantum dynamics
the threshold between delocalisation and self-trapping occurs at $k/{\cal E}_J=4/N$ \cite{milb,our}. The difference between the classical and quantum cases may be explained in terms
of the uncertainty relation between $z$ and $\phi$.  

It is apparent that the analysis of the classical system for non-zero external potential
only predicts one transition point,
which can be identified as a fixed point bifurcation. In constrast we have shown above that for the quantum dynamics
there are several points associated with minima and maxima of the energy gap, 
for which there is a qualitative transition in the dynamical behaviour.  
In Figs. (\ref{cjo},\ref{cgaps},\ref{ccross}) below, we show the dynamical behaviour predicted by the classical 
Hamiltonian using the same initial conditions and coupling parameters for the cases shown in 
Figs. (\ref{fig6},\ref{fig7},\ref{n40pq}) respectively. In particular, in figure \ref{ccross}
we present both, the classical and the quantum curves in a shorter 
time interval for a better comparison.

In most cases the results predicted by the quantum and classical analyses agree for sufficiently small 
time intervals, a feature also observed in \cite{milb}. In all cases however, there are stark differences in the 
dynamics over long time intervals. The irregular behaviour appearing below the transition 
coupling in 
Fig. \ref{fig6} does not occur in Fig. \ref{cjo}. The dynamical transitions that are shown in the 
last three columns of Fig. \ref{fig7} are not apparent from the classical results 
of Fig. \ref{cgaps}. 
And the collapse and revival of oscillations which is clear in the quantum 
dynamics (Fig.\ref{n40pq} and 
the solid curve in Fig.\ref{ccross}) is not seen in the 
classical case given by the dashed curve in Fig. \ref{ccross}. 
%%%%%%%%%
%\vspace{2cm}

%\vspace{1.0cm}
\begin{figure}
\begin{center}
\epsfig{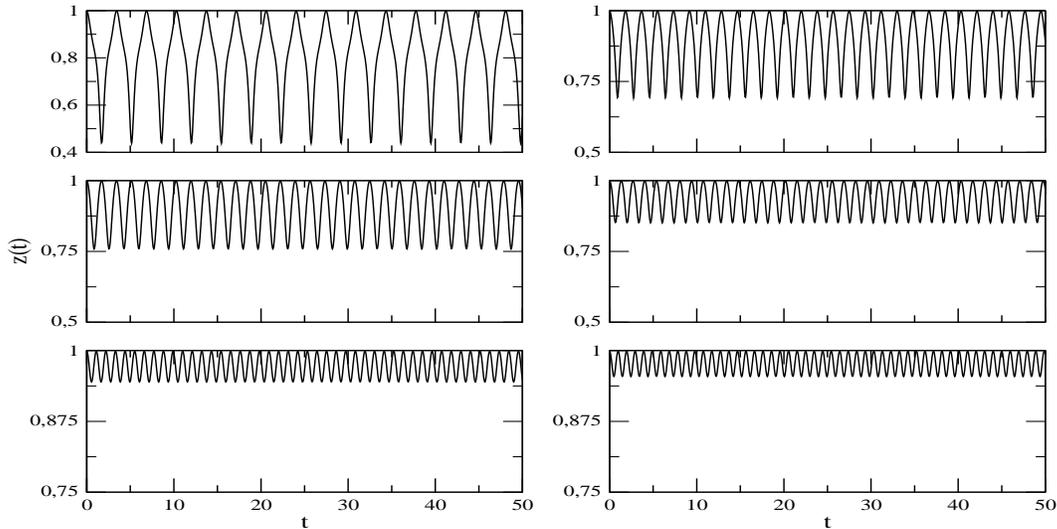}
\caption{Classical time evolution of the relative number of particles in the Josephson 
regime $k/{\cal E}_J=1$, for $N=50,100$. The initial condition is $z\simeq 1,\,\phi=0$. 
The central graph corresponds to the configuration where for the quantum model
the energy gap is a minimum, 
at $\Delta\mu_c/{\cal E}_J=24.5,\,49.5$. 
The other graphs represent points before and after this minimum point  
given by $\Delta\mu/{\cal E}_J= \Delta\mu_c/{\cal E}_J\mp5$.}
\label{cjo} 
\end{center}
\end{figure}
%%%%%%%%%%%%%%%%%%%%%
\vspace{1.0cm}
\begin{figure}
\begin{center}
\epsfig{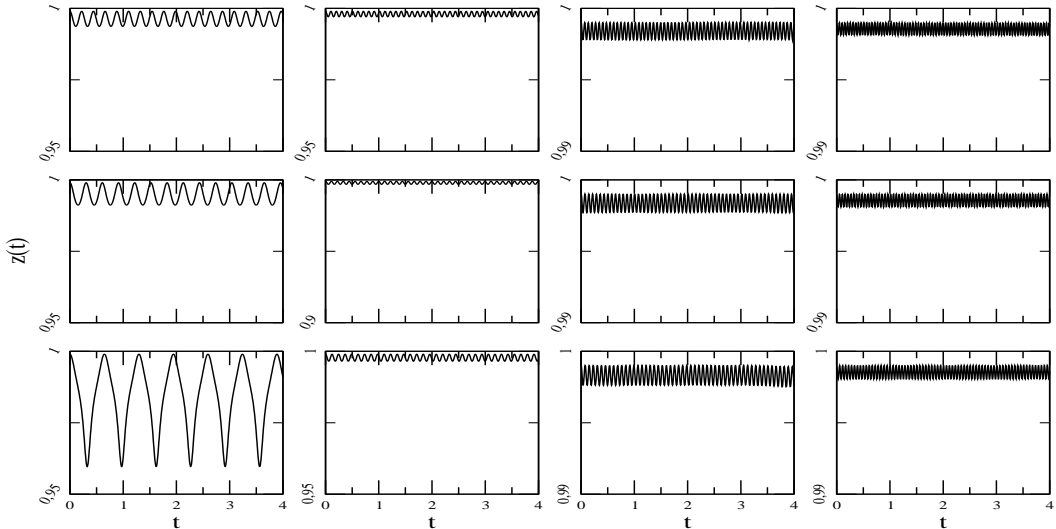}
\caption{Classical dynamics of the relative number of particles for $N=50$, $k/{\cal E}_J=60$ and with the initial 
condition $z\simeq 1,\,\phi=0$. The central graph, at each column, represents the point where for the quantum model the energy gap is minimal ($l=1,3$) or maximal ($l=2,4$), occurring at 
$ \Delta\mu_c/{\cal E}_J =1470,1440,1410,1380$ (from left to right).
In each case the graphs above and below  show dynamics of points 
around the minimum or maximum, 
chosen as  $\Delta\mu/{\cal E}_J =\Delta\mu_c/{\cal E}_J \mp 5$. }
\label{cgaps}
\end{center}
\end{figure}

%%%%%%%%%%%
\vspace{1.0cm}
\begin{figure}
\begin{center}
\epsfig{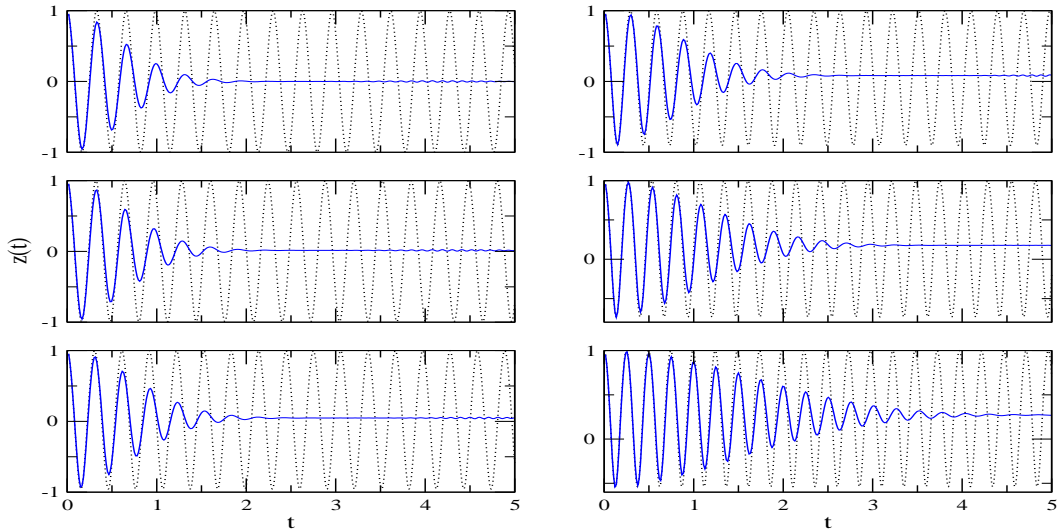}
\caption{Classical (dashed line) and quantum (solid line) evolution of the relative number of particles versus 
time for different values of the 
external potential (from the top to the bottom: $ \Delta\mu/k=0,1,3$ on the left and 
$ \Delta\mu/k=5,10,15$ on the right) in the regime between Josephson and Rabi 
(below the threshold coupling). Here we are fixing $N=40, {\cal E}_J=20$ and $k=1$.
The initial condition is $z\simeq 1,\,\phi=0$. }
\label{ccross}
\end{center}
\end{figure}

\section{Summary}
To summarise, we have studied the nature of the energy gap as a function of the 
external potential and its connection to the dynamical behaviour.  
We have shown that the energy gap function presents no minimum below 
the threshold coupling, while above this threshold point one minimum 
occurs close to the
Josephson regime and a set of minima and maxima appear in the Fock regime.
We found that the total number of these minima and maxima  
depends only on the number of particles.
An explicit expression for the position
of these minima and maxima was given.  
The connection between these energy gap extrema and the 
dynamics for the relative number of 
particles was investigated, and it was found that the extrema 
determine transition points separating different dynamical behaviours.
We have also compared these results with those that are obtained from a classical 
analysis of the Hamiltonian.

Finally, we refer to recent experimental work \cite{agfhco} where both delocalisation, with tunneling times
of the order 50 ms,  and self-trapping
have been confirmed in a single bosonic Josephson junction with a symmetric double-well potential using 
a Bose-Einstein condensate of $^{87}$Rb atoms. In this experimental 
setup both scenarios were observed by simply varying the total atomic population $N$. 
We note that from equation (\ref{rate2}) it can be seen that for fixed coupling parameters
$k,\,\Delta\mu$ and ${\cal E}_J$, the transition couplings $\Delta\mu_c$ can be tuned by also 
varying the total population $N$. Thus the results we have obtained 
above are relevant for the experimental system of \cite{agfhco} with an asymmetric potential.

\vspace{2.0cm}
\centerline{{\bf Acknowledgements}}
~\\
APT and AF thank CNPq-Conselho Nacional de Desenvolvimento
Cient\'{\i}fico e Tecnol\'ogico (a funding agency of the Brazilian Government)
for financial support. JL gratefully acknowledges funding from the Australian Research Council and The University of Queensland through a
Foundation Research Excellence Award. APT and AF would like to acknowledge GN Santos Filho for discussions.

%%%%%%%% %

\end{document}